\documentclass[useAMS,usenatbib]{mn2e}
\usepackage{graphicx}
\usepackage{lscape}
\usepackage{listings}

\title[T dwarfs in the GPS]{The discovery of the T8.5 dwarf UGPS J0521+3640 }
\author[Ben Burningham et al]{Ben Burningham$^{1}$\thanks{E-mail:
    B.Burningham@herts.ac.uk},  P.W. Lucas$^{1}$, S. K. Leggett$^{2}$, R. Smart$^{3}$,  D. Baker$^{1}$, D. J. Pinfield$^{1}$,
\newauthor
C. G. Tinney$^{4}$, D. Homeier$^{5}$,  F. Allard$^{6}$,  Z. H. Zhang$^{1}$,J. Gomes$^{1}$, A. C. Day-Jones$^{7}$,
\newauthor  
 H.R.A. Jones$^1$, G. Kov\'acs$^{8}$, N. Lodieu$^{9,10}$, F. Marocco$^{1}$, D. N. Murray$^{1}$, B. Sip\H{o}cz$^{1}$\\
$^{1}$ Centre for Astrophysics Research, Science and Technology Research Institute, University of Hertfordshire, Hatfield AL10 9AB \\
$^{2}$ Gemini Observatory, 670 N. A'ohoku Place, Hilo, HI 96720, USA \\
$^{3}$ Istituto Nazionale di Astrofisica, Osservatorio Astronomico di Torino, Strada Osservatrio 20, 10025 Pino Torinese, Italy \\
$^{4}$ School of Physics, University of New South Wales, 2052. Australia\\ 
$^{5}$ Institut fur Astrophysik, Georg-August-Universitat, Friedrich-Hund-Platz 1, 37077 Gottingen, Germany \\
$^{6}$ C.R.A.L. (UMR 5574 CNRS), Ecole Normale Superieure, 69364 Lyon Cedex 07, France \\
$^{7}$ Universidad de Chile,Camino el Observatorio \# 1515, Santiago, Chile, Casilla 36-D\\
$^{8}$ Institute of Astronomy, Madingley Road, Cambridge CB3 0HA, UK\\
$^{9}$ Instituto de Astrof\'isica de Canarias (IAC), Calle V\'ia L\'actea s/n, E-38200 La Laguna, Tenerife, Spain\\
$^{10}$ Departamento de Astrof\'isica, Universidad de La Laguna (ULL), E-38205 La Laguna, Tenerife, Spain\\
}

\begin{document}
%
%
%
%


\def\aj{\rm{AJ}}                   
\def\araa{\rm{ARA\&A}}             
\def\apj{\rm{ApJ}}                 
\def\apjl{\rm{ApJ}}                
\def\apjs{\rm{ApJS}}               
\def\ao{\rm{Appl.~Opt.}}           
\def\apss{\rm{Ap\&SS}}             
\def\aap{\rm{A\&A}}                
\def\aapr{\rm{A\&A~Rev.}}          
\def\aaps{\rm{A\&AS}}              
\def\azh{\rm{AZh}}                 
\def\baas{\rm{BAAS}}               
\def\jrasc{\rm{JRASC}}             
\def\memras{\rm{MmRAS}}            
\def\mnras{\rm{MNRAS}}             
\def\pra{\rm{Phys.~Rev.~A}}        
\def\prb{\rm{Phys.~Rev.~B}}        
\def\prc{\rm{Phys.~Rev.~C}}        
\def\prd{\rm{Phys.~Rev.~D}}        
\def\pre{\rm{Phys.~Rev.~E}}        
\def\prl{\rm{Phys.~Rev.~Lett.}}    
\def\pasp{\rm{PASP}}               
\def\pasj{\rm{PASJ}}               
\def\qjras{\rm{QJRAS}}             
\def\skytel{\rm{S\&T}}             
\def\solphys{\rm{Sol.~Phys.}}      
\def\sovast{\rm{Soviet~Ast.}}      
\def\ssr{\rm{Space~Sci.~Rev.}}     
\def\zap{\rm{ZAp}}                 
\def\nat{\rm{Nature}}              
\def\iaucirc{\rm{IAU~Circ.}}       
\def\aplett{\rm{Astrophys.~Lett.}} 
\def\apspr{\rm{Astrophys.~Space~Phys.~Res.}}
\def\bain{\rm{Bull.~Astron.~Inst.~Netherlands}} 
\def\fcp{\rm{Fund.~Cosmic~Phys.}}  
\def\gca{\rm{Geochim.~Cosmochim.~Acta}}   
\def\grl{\rm{Geophys.~Res.~Lett.}} 
\def\jcp{\rm{J.~Chem.~Phys.}}      
\def\jgr{\rm{J.~Geophys.~Res.}}    
\def\jqsrt{\rm{J.~Quant.~Spec.~Radiat.~Transf.}}
\def\memsai{\rm{Mem.~Soc.~Astron.~Italiana}}
\def\nphysa{\rm{Nucl.~Phys.~A}}   
\def\physrep{\rm{Phys.~Rep.}}   
\def\physscr{\rm{Phys.~Scr}}   
\def\planss{\rm{Planet.~Space~Sci.}}   
\def\procspie{\rm{Proc.~SPIE}}   

\let\astap=\aap
\let\apjlett=\apjl
\let\apjsupp=\apjs
\let\applopt=\ao

\maketitle

\begin{abstract}
We have carried out a search for late-type T dwarfs in the UKIDSS Galactic Plane Survey 6$^{th}$ Data Release. The search yielded two persuasive candidates, both of which have been confirmed as T dwarfs. 
The brightest, UGPS~J0521+3640 has been assigned the spectral type T8.5 and appears to lie at a distance of 7--9~pc. The fainter of the two, UGPS~J0652+0324, is classified as a T5.5 dwarf, and lies at an estimated distance of 28--37~pc.
Warm-{\it Spitzer} observations in IRAC channels 1 and 2, taken as part of the GLIMPSE360 Legacy Survey, are available for UGPS~J0521+3640 and we used these data with the near-infrared spectroscopy to estimate its properties. We find best fitting solar metallicity  BT-Settl models for  $T_{\rm eff} = 600$K and 650K and $\log g = 4.5$ and 5.0. These parameters suggest a mass of  between 14 and 32~M$_{Jup}$ for an age between 1 and 5~Gyr.
The proximity of this very cool T dwarf, and its location in the Galactic plane makes it an ideal candidate for high resolution adaptive optics imaging to search for cool companions.

\end{abstract}


\section{Introduction}
\label{sec:intro}
The growing sample of very cool T dwarfs \citep[e.g. ][]{warren07,ben08,delorme08,ben09,delorme10,goldman2010,lucas2010,mainzer2011} is providing a crucial test bed for atmospheric model grids that span the stellar, substellar and planetary regime \citep{marley02,sm08,allard2010}.
A number of recent discoveries \citep[e.g. ][]{folkes2007, artigau2010} including the very cool and nearby T~dwarf UGPS~J0722-0540 \citep{lucas2010} demonstrate that despite the technical challenges associated with identifying rare red objects in the Galactic plane, it is a feasible region in which to identify such targets.
Nearby T dwarfs in the Galactic plane offer a number of advantages that make them ideal for detailed study. 
In addition to their relative brightness compared to more distant sources, the greater likelihood of proximity to stars bright enough to act as natural guide stars or tip-tilt stars for laser guide star observations make cool brown dwarfs in the Galactic plane ideal for high-resolution imaging campaigns aimed at identifying cool substellar neighbours and dynamical benchmark systems \citep[e.g. ][]{liu08,dupuy09a,dupuy09b}.

In this Letter we present the results of a new search of the UKIRT Infrared Deep Sky Survey \citep[UKIDSS;  ][]{ukidss} Galactic Plane Survey \citep[GPS; ][]{gps} 6$^{th}$ Data Release for nearby late-type T dwarfs which allows for candidates in the colour range $-0.1 <H-K < +0.1$ that were excluded by the search reported in \citet{lucas2010} .

\section{Candidate selection}
\label{sec:sel}

Late T dwarfs have a fairly wide spread of near infrared colours that
includes objects with $(J-H) > 0$ or $(H-K) > 0$. A quite
restrictive selection ($(J-H) < -0.2$, $(H-K) < -0.1$) was employed to 
detect UGPS~0722-0540 in order to minimise the number 
of false positive candidates with colours similar to normal stars.
Inspection of the two colour diagrams in figure 9 of \citet{ben10b}
shows that T8.5 and T9 dwarfs typically have $(H-K) \approx 0$ and 
$(J-H) < -0.2$. A query of the 6$^{th}$ UKIDSS GPS Data Release at 
WFCAM Science Archive \citep{wsa} showed that the number of candidates in the gpsSource
catalogue is increased only modestly if the $(H-K)$ colour selection 
is relaxed to $(H-K) < +0.1$, provided that all the other selection
criteria employed in \citet{lucas2010} remain in place. The number
of candidates requiring visual inspection rises much more rapidly if this 
colour cut is relaxed further, or if the $(J-H)$ selection is relaxed.

We therefore searched the colour space $-0.1< (H-K)<+0.1$, 
$(J-H) < -0.2$ for additional late T dwarfs. We note that this
search would miss unusually red objects such as the T8.5 dwarfs ULAS~J1302+1308, 
and Ross 458C which have  $(H-K) = 0.32$ and 0.11 respectively. 
In other respects the search was identical to that carried described in  \citet{lucas2010}, to which the reader is referred for an explanation of the logic behind the selection criteria.

 This selection yielded 12 candidates. Two of these, UGPS~J0521+3640
and UGPS~J0652+0324, passed visual inspection of the FITS images and have no
optical counterpart in either the IPHAS survey \citep{drew2005} or the 
POSS USNO-B1.0 archive. These two were therefore selected for spectroscopic
observation. An additional search was run that removed the requirement for the coordinate shifts between the three passbands to be smaller than 0.3$\arcsec$ and included all sources with $H-K<0.1$ mag that satisfy the other  criteria above. This search produced only 3 additional candidates, all of which were identified as image defects upon inspection of the FITS images.

\section{Observations}
\label{sec:obs}
The near-infrared discovery images of UGPS~J0521+3640 were obtained on 25$^{th}$ October 2007, whilst UGPS~J0652+0324 was observed on 29$^{th}$ October 2007. 
Follow-up imaging to measure proper motions was obtained use the Near Infrared Camera Spectrometer (NICS) on the National Telescope Galileo (TNG) on La Palma during the nights of 8$^{th}$ February and 28$^{th}$ January 2011 respectively. The data were processed using the standard NICS pipeline\footnote{http://www.tng.iac.es/news/2002/09/10/snap/}. 
Additional WFCAM \citep{wfcam} imaging for both targets was obtained on 21$^{st}$ November 2010 and further images were obtained for  UGPS~J0521+3640 on 2$^{nd}$, 18$^{th}$ and 19$^{th}$ February 2011. 
The WFCAM data were processed using the standard pipeline \citep{irwin04}. The $x,y$ coordinates of the targets in all frames were converted to the standard coordinate system of the discovery frames using a simple linear model. 
The relative proper motion for all objects were found from linear fits
to the standard coordinates at the different epochs. 
We were unable to determine a self-consistent motion for UGPS~J0652+0324 from the 3 epochs of data available.
A correction to an
absolute system was estimated from the median difference between measured
relative proper motions and published values of PPMXL \citep{ppmxl} objects in the field.
The derived proper motion for UGPS J0521+3640 was corrected for an assumed
parallax of 125mas. The UKIDSS coordinates and proper motions are given in Table~\ref{tab:pm}.  
The Simbad and SuperCOSMOS \citep{supercos1} databases were searched for targets to which UGPS~J0521+3640 may be a common proper motion companion, but no viable candidates were identified. 

\begin{table*}
\begin{tabular}{|c c c c c c c|}
Object & $\alpha$ & $\delta$ & Epoch & Equinox & $\mu_{\alpha \cos \delta}$ / mas yr$^{-1}$  & $\mu_{\delta}$  / mas yr$^{-1}$ \\
\hline
 UGPS J0521+3640 &  05:21:27.27 &  +36:40:48.6 & 2007.82 & J2000 & $513 \pm 10$ & $- 1507 \pm 10$ \\
UGPS J0652+0324 &  06:52:27.71 & +03:24:31.0 & 2007.83 & J2000 & & \\
\hline
 \end{tabular}
 \caption{UKIDSS discovery coordinates for the targets and the proper motions for UGPS J0521+3640.}
\label{tab:pm}
\end{table*}

 Warm-Spitzer IRAC [3.6] and [4.5] imaging of the region around UGPS J0521+3640 was obtained as part of the GLIMPSE360 legacy program\footnote{ http://www.astro.wisc.edu/sirtf/glimpse360/}. 
 The [3.6] data were obtained on 31$^{st}$ October 2009, with AORkey 32886784; the [4.5] data on 30$^{th}$ October 2009, with AORkey 32912896. 
 The frame time in both cases was 12s.
 The post-basic-calibrated-data (pbcd) mosaics generated by version 18.18.0 of the Spitzer pipeline were used to obtain aperture photometry. The photometry was derived using a 3.6$\arcsec$-radius aperture, and the aperture correction was taken from the IRAC handbook\footnote{http://ssc.spitzer.caltech.edu/irac/dh/}. 
 UGPS~J0521+3640 is currently close enough to a bright star (36$\arcsec$ North-NorthWest) that care had to be taken with subtraction of the sky background. The choice of sky background was investigated and found to lead to an uncertainty of the same size as that indicated by the noise statistics in the pbcd uncertainty image.  The error quoted in Table~\ref{tab:photo} is the result of adding these random errors in quadrature to the 3\% error that accounts for systematic effects due to calibration uncertainties and pipeline dependencies.

\begin{table*}
\begin{tabular}{|c c c c c c c c c|}
Object & $J_{UKIDSS}$ & $H_{UKIDSS}$ & $K_{UKIDSS}$ & [3.6] & [4.5] & $J-H$ & $H-K$ & $H - $[4.5] \\
\hline
UGPS J0521+3640 &  $16.94 \pm 0.02$ & $17.28 \pm 0.04$ & $17.32 \pm 0.09$ & $15.30 \pm 0.06$   & $13.36 \pm 0.03$ & $-0.34 \pm 0.04$ & $-0.04 \pm 0.1$ & $3.92 \pm 0.05$ \\
UGPS J0652+0324 & $17.24 \pm 0.02$  & $17.50 \pm 0.04$ & $17.52 \pm 0.09$ & & & $-0.26 \pm 0.04$ & $-0.02 \pm 0.1$ & \\
\hline
\end{tabular}

\caption{Summary of photometry for UGPS~J0521+3640 and UGPS~J0652+0324. 
\label{tab:photo}}

\end{table*}

Near-infrared spectroscopy of UGPS~J0521+3640 and UGPS~J0652+0324 was obtained using the Gemini Near InfraRed Spectrograph \citep[GNIRS; ][]{elias2006} mounted on the Gemini-North telescope.  
UGPS~J0521+3640 was observed on the night of 30$^{th}$ December 2010 with a total integration time of 24 minutes. UGPS~J0652+0324 was observed on the night of 26$^{th}$ December 2010 with a total integration time of 40 minutes.
The targets were observed in cross-dispersed mode capturing the full 0.8--2.5$\micron$ region with a 0.68$\arcsec$ slit delivering a resolving power of R$\sim 1200$.  The data were reduced using GNIRS routines in the Gemini {\sc IRAF} package \citep{cooke2005}, using a nearby F star in each case for telluric correction. 
The telluric standard spectrum was divided by a blackbody spectrum of an appropriate $T_{\rm eff}$ after removing hydrogen lines by interpolating the local continuum. The rectified standard spectrum was then used to correct for telluric absorption and to provide relative flux calibration. The overlap regions between the orders in the $Y$,$J$ and $H$ bands agreed well suggesting that the relative flux of the orders is well calibrated. 
The resulting $YJHK$ spectra  are shown in Figure~\ref{fig:jhkspec}.

\section{Analysis}
\label{sec:anal}

\subsection{Spectral types}
\label{subsec:sptype}

\begin{figure*}
\includegraphics[height=500pt, angle=90]{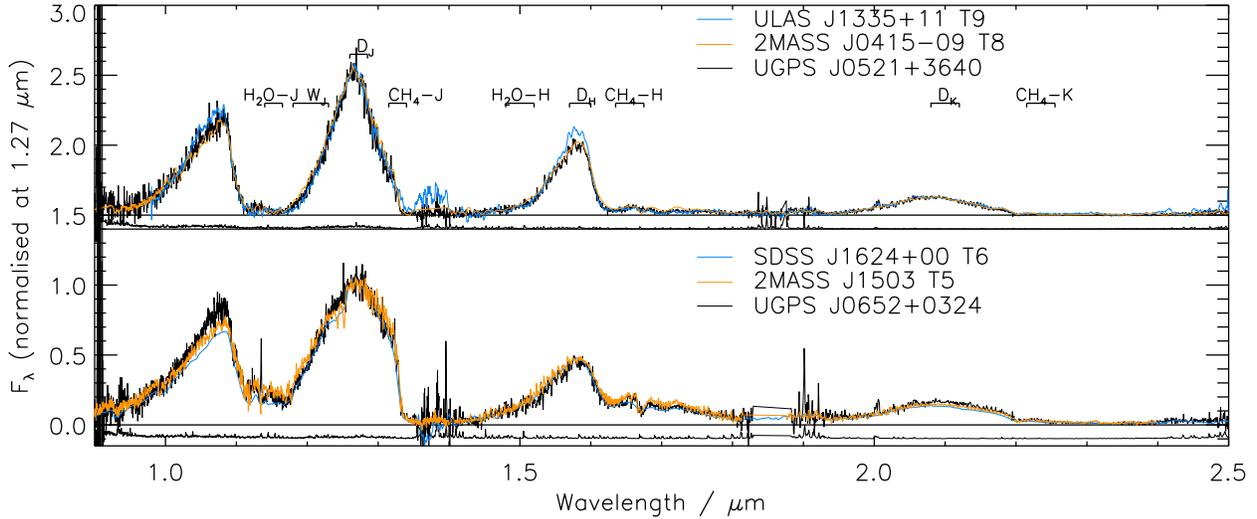}
\caption{GNIRS spectra for the T dwarfs presented here compared to T dwarf spectral type templates taken from \citet{burgasser06} and \citet{ben08}.}
\label{fig:jhkspec}
\end{figure*}

We have classified the two T dwarfs presented here following the scheme of \citet{burgasser06}, with the extension to T9 as discussed in \citet{ben08} . Spectral typing index ratios for the targets presented here are given in Table~\ref{tab:ratios}. In Figure~\ref{fig:jhkspec} we compare our targets to appropriate spectral type templates. Both the spectral typing index ratios and the template comparison suggests a type of T5.5 $\pm 0.5$ for UGPS~J0652+0324.
The spectrum of UGPS~J0521+3640 appears to be intermediate between the T8 and T9 spectral type templates 2MASS~J0415-09 and ULAS~J1335+1130. 
Its $W_J$ index lies towards the upper end of the range defined for T9 in \citet{ben08}, and is most similar to the value found for the T8.5 dwarf Wolf~940B, whilst its H$_2$O--H index has a value towards the lower range of the T8 bracket. 
The other indices that were defined by \citet{burgasser06} are degenerate for types T8 and T9. 
The NH$_3$--H index, defined by \citet{delorme08}, does not easily map onto the T dwarf classification system for earlier types since it is degenerate for types $\leq$T8 \citep{ben08}. However, its value for UGPS~J0521+3640 ($0.597 \pm 0.003$) is intermediate between the values found for T7--T8 dwarfs and those typically seen for T9 dwarfs, suggesting that a later-than T8 classification is justified.
Based on these considerations we assign a spectral type of T8.5 $\pm 0.5$ for UGPS~J0521+3640.

Applying the spectral type vs M$_J$ relations of  \citet{marocco2010}, we thus estimate distances of $8.2^{+1.2}_{-1.0}$~pc and $32^{+5}_{-4}$~pc for these objects. 
The uncertainties on these distances are derived from the uncertainties in the polynomial coefficients quoted by  \citet{marocco2010}. 
We have not included the uncertainty due to our $\pm 0.5$ subtype spectral typing precision. 
For the coolest dwarfs, the spectral type -  M$_J$ relation is sufficiently steep that a half subtype uncertainty can have a large effect on the inferred distance. 
In the case of UGPS~J0521+3640 we find that a T8--T9 range in spectral type corresponds to distance bracket of 5.9 -- 11.4~pc.

\begin{table*}
\begin{tabular}{| c c c c c c c c |}
Object & H$_2$O--J & CH$_4$--J & $W_J$ & H$_2$O--H & NH$_3$--H & CH$_4$--H &  CH$_4$--K \\
\hline
U0521 &  $0.028 \pm 0.001$  & $0.147 \pm 0.001$ & $0.269 \pm 0.001$ &   $0.151 \pm 0.002$ & $0.597 \pm 0.003$ & $0.103 \pm 0.001$ & $0.036 \pm 0.002$  \\
 & $\geq$T8 &  $\geq$T8  & T9  &  T8 & $>$T8 &  $\geq$T8 & $\geq$ T7 \\
U0652 & $0.199 \pm 0.002$& $0.361 \pm 0.001$ & $0.491 \pm 0.002$ & $0.325 \pm 0.003$ & $0.671 \pm 0.001$  & $0.362 \pm 0.002$ & $0.182 \pm 0.002$ \\
 &  T5  &  T5/6 & $<$T7 & T5/6 & $<$T8 & T5/6 &  T5/6 \\
\hline
\end{tabular}
\caption{Spectral typing index ratios for UGPS~J0521+3640 and UGPS~J0652+0324 as defined by \citet{burgasser06} and extended to T8+ by \citet{ben08}. The $W_J$ ratio was defined by \citet{warren07}, whilst the NH$_{3}$--H ratio was defined by \citet{delorme08}.
\label{tab:ratios}}

\end{table*}



\subsection{Properties of UGPS J0521+3640}
\label{subsec:props}

\begin{figure}
\includegraphics[height=250pt, angle=90]{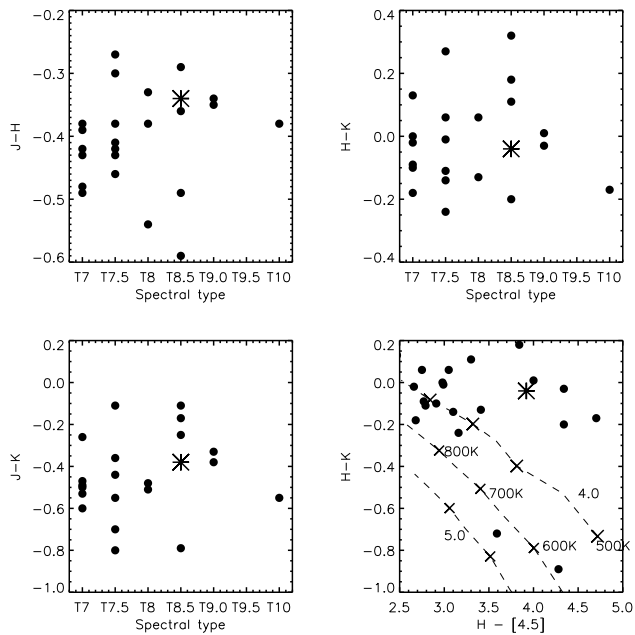}
\caption{Colour versus spectral type plots for T7--T10 dwarfs with MKO photometry, along with a $H-K$ / $H-$[4.5] colour-colour plot shown in the bottom right panel. Synthetic colours for  solar metallicity \citet{saumon06,saumon07} are shown as solid lines with crosses indicating 100K intervals. The lower left line corresponds to $\log g = 4.0$, the middle line $\log g = 4.5$ and the upper right line $\log g = 5.0$. .  UGPS~J0521+3640 is shown with an asterisk. 
}
\label{fig:colplot}
\end{figure}

In Figure~\ref{fig:colplot} we show the near-infrared colour versus spectral type plots and a $H-K$ / $H-$[4.5] for T7--T10 dwarfs, placing UGPS~J0521+3640 in context within the wider sample. 
Both $H-K$ and $J-K$ colours (the latter often discussed in terms of the $J/K$ ratio) have been suggested as useful metallicity and gravity diagnostics. 
Poor agreement between the model atmospheres and observations of the flux ratio between the $K$ band and the $J$ and $H$ bands \citep[e.g.][]{ben09,ben11} preclude their use for making absolute estimates for gravity or metallicity, but they allow useful relative comparisons between objects. 
The location of UGPS~J0521+3640 on the plots in Figure~\ref{fig:colplot} suggests that its metallicity and gravity are within the range of values seen for other T8--T9 dwarfs.

We have estimated the properties of UGPS~J0521+3640 by comparison with the latest BT Settl model grid \citep{btsettlCS16} for solar metallicity and a reasonable $T_{\rm eff}$ and gravity range (based on experience of objects of similar spectral type).
We have used the same method as described in Burningham et al (2011), which is an adaptation of the \citet{cushing08} method for fitting model spectra that incorporates flux information from photometric data points (e.g our {\it warm}-Spitzer photometry). The lack of a known parallax for our target means that we can only consider the case of the freely scaled model spectra, and cannot use the bolometric luminosity to better constrain the properties.
Prior to fitting the models  the target spectrum was placed on an absolute flux scale using the $J$ band photometry. Scaling by the noisier $Y$, $H$ and $K$ photometry provided consistent results.

We find that the goodness of fit statistic, $G$, is minimised for the cases where $T_{\rm eff} = 600$K and $\log g = 4.5$ and $T_{\rm eff} = 650$K and $\log g = 5.0$.  
These values for $T_{\rm eff}$ are broadly consistent with the location of UGPS~J0521+3640 on the $H-K$ / $H-$[4.5] plot shown in Figure~\ref{fig:colplot}, although, as has been noted previously, the models predict colours that are too blue in $H-K$.
These properties correspond to a mass of 14 -- 32~M$_{Jup}$, and a radius of approximately 1.0 -- 0.8~R$_{Jup}$ for ages of 1--5~Gyr on the COND evolutionary model grid \citep{baraffe03}, although an age of 10 Gyr and a mass of 35~M$_{Jup}$ would be implied by a moderately higher surface gravity that is not resolved by our model grid.
These best fits are found for scaling factors that correspond to distance to radius ratios of $d / R = 8.00$ and $9.78~{\rm pc} / R_{Jup}$ respectively.  
These are consistent with the photometric distances determined in Section~\ref{subsec:sptype} , with implied distances of 8~pc and 7.8~pc for the inferred radii of each case. 
This suggests that the distance derived for the T8.5 classification in Section~\ref{subsec:sptype}, without the inclusion of the $\pm 0.5$ subtype uncertainty, of $8.2^{+1.2}_{-1.0}$~pc is a reasonable best-estimate for the distance to UGPS~J0521+3640.

At this distance, the proper motion of UGPS~J0521+3640 ($\mu_{\alpha \cos \delta} = 513$~mas yr$^{-1}$, $\mu_{\delta} = -1507$~mas yr$^{-1}$) corresponds to a tangential velocity, $V_{tan} = 60$~kms$^{-1}$. This is at the upper end of the distribution for T~dwarfs reported by \citet{vrba04}, implying that this source is unlikely to be very young (e.g. $< 1$Gyr), and supporting our estimated age range of 1--5~Gyr.

\section{Summary}
\label{sec:summ}
We have searched the 6$^{th}$ data release of the UKIDSS GPS for late type T~dwarfs and have identified two such objects with types T5.5 and T8.5. 
The T8.5 dwarf, UGPS~J0521+3640, appears to lie at distance of less than 10~pc, and as such is an ideal candidate for high resolution imaging searches for cool companions and detailed characterisation. 
We have used archival warm-{\it Spitzer} observations and the latest BT Settl model grid to estimate that UGPS~J0521+3640 is cooler than 700K, with best fitting models at $T_{\rm eff} =600$ and 650K.

\section*{Acknowledgements}

We thank our colleagues in the GLIMPSE360 team for providing
this deep panoramic Legacy dataset, which complements the UKIDSS
GPS excellently. We thank Barbara Whitney, Brian Babler and Robert
Benjamin for a helpful discussion of the Spitzer photometry.
Based on observations made under project A22TAC$\_$96 on the Italian Telescopio Nazionale Galileo (TNG) operated on the island of La Palma by the Fundaci—n Galileo Galilei of the INAF (Istituto Nazionale di Astrofisica) at the Spanish Observatorio del Roque de los Muchachos of the Instituto de Astrofisica de Canarias. Based on observations obtained via program GN-2010B-Q-41 at the Gemini Observatory,
which is operated by the
Association of Universities for Research in Astronomy, Inc., under a cooperative agreement
with the NSF on behalf of the Gemini partnership: the National Science Foundation (United
States), the Science and Technology Facilities Council (United Kingdom), the
National Research Council (Canada), CONICYT (Chile), the Australian Research Council (Australia),
Minist\'{e}rio da Ci\^{e}ncia e Tecnologia (Brazil)
and Ministerio de Ciencia, Tecnolog\'{i}a e Innovaci\'{o}n Productiva (Argentina).
SKL's research is supported by the Gemini Observatory.
CGT is supported by ARC grant DP0774000.
ADJ is supported by a Fondecyt postdoctorado fellowship, under project number 3100098.
This research has made use of the SIMBAD database,
operated at CDS, Strasbourg, France, and has benefited from the SpeX
Prism Spectral Libraries, maintained by Adam Burgasser at
http://www.browndwarfs.org/spexprism.
JG, GK and BS are supported by RoPACS, a Marie Curie Initial
Training Network funded by the European CommissionÕs Seventh Framework
Programme. NL acknowledges funding from the Spanish Ministry of Science and Innovation through the Ram\'on y Cajal fellowship number 08-303-01-02 and the project number AYA2010-19136.
The authors wish to recognise and acknowledge the very significant cultural role and reverence that the summit of Mauna Kea has always had within the indigenous Hawaiian community.  We are most fortunate to have the opportunity to conduct observations from this mountain.
\bibliographystyle{mn2e}
\bibliography{refs}

\end{document}